\documentclass[12pt,a4paper]{article}
\usepackage{graphicx}
\usepackage{amssymb}
\usepackage{cite}

\begin{document}

\title{The Fate of Ernst Ising and the Fate of his Model}
\author{\em Thomas Ising$^{\,a}$, Reinhard Folk$^{\,b}$, Ralph Kenna$^{\,c,a}$,\\
\em Bertrand Berche$^{\,d,a}$, Yurij Holovatch$^{\,e,a}$\\
\\{\scriptsize
$^a$${\mathbb L}^4$ Collaboration \& Doctoral College for the
Statistical Physics of
Complex Systems,}\\
{\scriptsize  Leipzig-Lorraine-Lviv-Coventry}\\
{\scriptsize $^b$ Institute for Theoretical Physics, Johannes Kepler
University Linz, A-4040, Linz, Austria}\\
{\scriptsize
$^c$Applied Mathematics Research Centre, Coventry University, Coventry, CV1 5FB, United Kingdom}\\
{\scriptsize
$^d$Statistical Physics Group,
Laboratoire de Physique et Chimie Th\'eoriques,}\\
{\scriptsize Universit\'e de
Lorraine, F-54506 Vand\oe uvre-les-Nancy Cedex, France}\\
{\scriptsize $^e$Institute for Condensed Matter Physics, National
Acad. Sci. of Ukraine, UA--79011 Lviv, Ukraine}  }
\date{June 6, 2017}

\maketitle
\begin{abstract}
On this, the occasion of the 20th anniversary of the ``Ising Lectures'' in Lviv
(Ukraine), we give some personal reflections about the famous
model that was suggested by Wilhelm Lenz for ferromagnetism in 1920 and solved
in one dimension by his PhD student, Ernst Ising, in 1924. That work of Lenz and Ising
marked the start of a scientific direction that, over nearly 100 years, delivered extraordinary
successes in explaining collective behaviour in a vast variety of systems, both within and
beyond the natural sciences. The broadness of the appeal of the Ising model is reflected in the variety of talks presented at the
Ising lectures ({\tt http://www.icmp.lviv.ua/ising/}) over the past two decades
but requires that we restrict this  report to a small selection of topics.
The paper starts with some personal memoirs of Thomas Ising (Ernst's son).
We then discuss the history of the model, exact solutions, experimental
realisations, and its extension to other fields.

\end{abstract}

\section{Introduction}
\label{intro}

In seeking to explain a particular phenomenon in physics, namely the
onset of ferromagnetism, Wilhelm Lenz proposed a model that was
solved in one dimension by his PhD student, Ernst Ising in 1924.
This event marked the start of a process that, over nearly 100
years, delivered  tremendous and multiple successes in explaining
collective behaviour in a vast variety of systems, including many
beyond the natural sciences.

\begin{figure}
\centerline{\includegraphics[height=9.0cm]{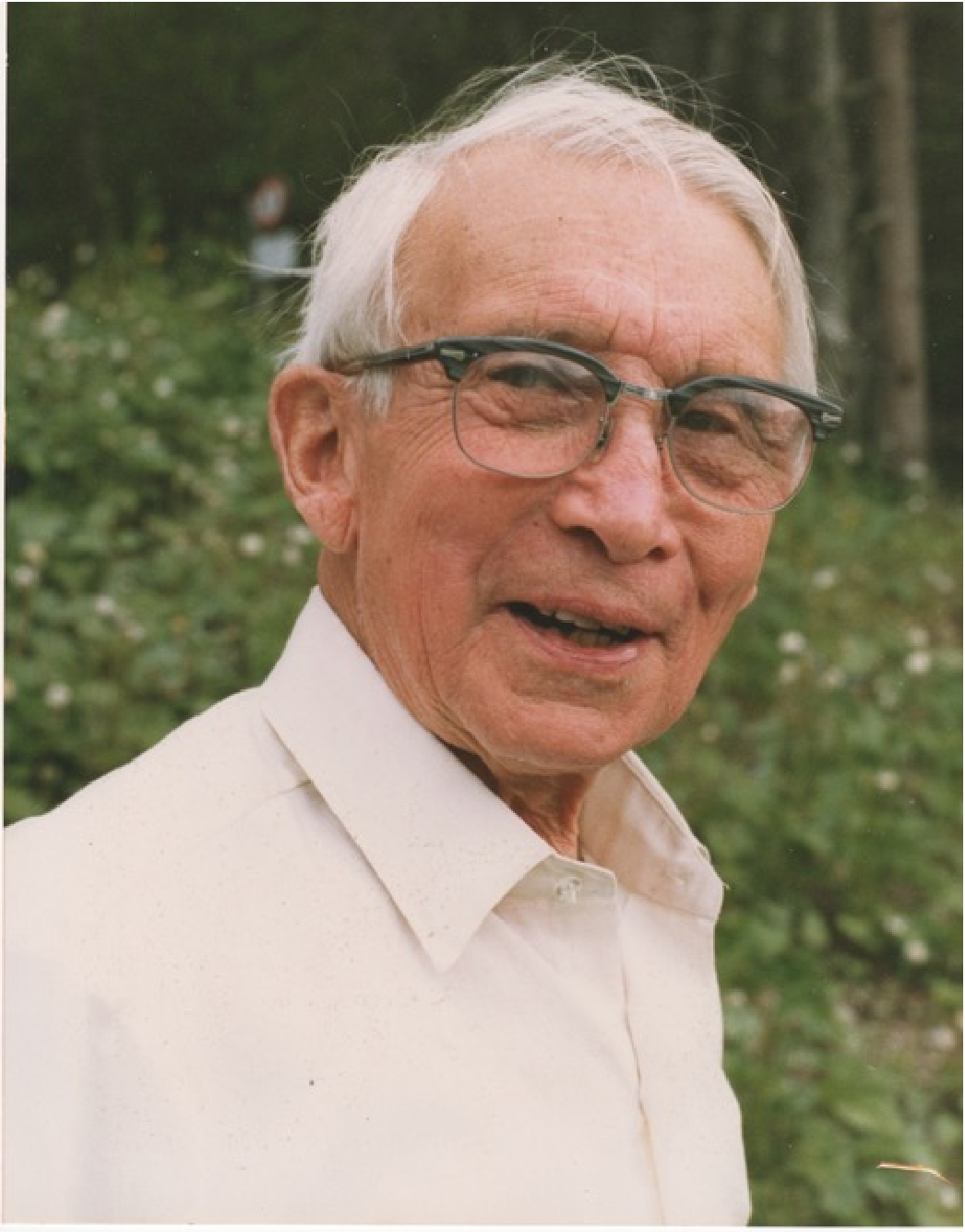}}
\caption{\label{Ising1987}Ernst (Ernest) Ising (May 10, 1900 in
Cologne, Germany - May 11, 1998 in Peoria, USA). Photo taken in
1987.}
\end{figure}

When Ernst Ising started his work on the phenomenon of
ferromagnetism, the nature of the microscopic, inter-atomic
interactions was not yet understood. Indeed, it was not at all clear
how a macroscopic magnetization could be generated by the
interactions between elementary magnets. It was already known that
magnetism is a quantum phenomenon but quantum theory was at a stage
where the classical Bohr-Sommerfeld model was in disagreement with
experiments. Ernst Ising succeeded in answering the question in the
specific context of Lenz's suggestion for a linear chain. At the
same time, Wolfgang Pauli, who was also at Lenz's institute,
contributed to quantum mechanics by suggesting that the electron
possesses a two valued non-classical magnetic moment. While these
were important steps towards answering the above questions, the full
picture had to wait for further new ideas.

Nowadays literature based on  the Ising model is abundant and there
are several good reviews that report the history of the model
\cite{brush,Niss,kobe} and its applications in different fields of
science \cite{Castellano09,Galam12,Sornette14}. About 800 papers on
the Ising model are published every year \cite{kobe}$^{(b)}$. It has
found applications in a range of different circumstances such as
tumor modelling \cite{torquato11}, seismic-hazard assessment
\cite{jimenz07} and sonification of science (instead of
visualization) \cite{vogt07}. Quite recently a universal simulator
modelling   spin models has been found \cite{cuevas16}. Such broad
impact of the Ising model is also reflected in the contributions to
the ``Ising lectures'' --  an annual workshop in Lviv that in 2017
celebrates   its 20th anniversary \cite{Ising_lectures},  the
occasion for which this paper is written. The broadness of the
appeal of the Ising model requires that we restrict our report to a
small selection of topics. In particular, in what follows we discuss
the history of the model and its formulation as we now know it
(Section \ref{hist});
 exact solutions (Section \ref{Exact1});
 experimental realisations (Section \ref{experiment});
and its extension to other fields (Section \ref{exot}). We start
with the personal memoirs of Thomas Ising (Ernst's son) in the next
section.

\section{My Father - Ernest Ising}
\label{personal}

My father was a wonderful person who was in love with life. He
thoroughly enjoyed teaching:  ``I got some of the students in the
front row wet with my experiment." He often stated that no class was
complete unless his students had laughed with him.

When I first got to know him physics was something far away. He was
only interested in keeping himself and our family alive in the
middle of WWII.
Starting with 1933, there were really only twelve very bad years for him.

He was born Ernst Ising, at the beginning of the 20th century on May
10th in K\"oln (or Cologne) near the cathedral. His mother, Thekla
Loewe, came from a very successful Jewish merchant family in
Duisburg. His father, Gustav, grew up in the small town of Rietberg
in rural Westphalia,  as son of the local blacksmith. We do not know
how or when Thekla and Gustav met, but Kaufhaus Loewe had quite a
few male and female employees. Thekla, my grandmother, spoke of the
table usually being set for about forty people. Gustav and his
brother (?) Bernard ran a very successful upscale women's clothing
store in Bochum until the Hitler years. By the time his sister,
Charlotte (Lotte), was born in 1904, his parents had a wonderful
home in a wealthier part of Bochum. Their home became a stopping
point for many artistes of the period.  It included a two-story
stained glass stairwell by Johan Thorn Prikker. This unfortunately
did not survive the war.

My father liked acting and had a stage in the basement where he and
his friend, Heinz Wildhagen put on plays. Heinz spent his life as an
actor and theater owner.  Later the actor Willie Busch became my
father's dictation coach.

After completing Gymnasium in 1918 he spent a few compulsory months
as a soldier near the end of WWI. On the day that the war ended he
was on a ladder hanging a banner. He said he looked around and
everyone was gone! The war was over and they had all left.

In 1919 he started at the university in G\"ottingen majoring in math
and physics. Later he was at the University in Bonn.

In graduate school at Hamburg University in 1922 he came under the
tutelage of Professor Wilhelm Lenz who suggested a doctoral thesis
in ferromagnetism following up on his paper of 1920.  The thesis was
completed in 1924. One of his fellow students was Wolfgang Pauli.
Also at this time his sister, Lotte, married Hermann Busch (Willie's
brother) of the famous Busch family.

After receiving his PhD he went to work in the patent office of the
AEG or the Allgemeine Elektrizit\"atsgesellschaft (General Electric)
in Berlin. While he enjoyed the work, he knew that he preferred
teaching. During this time he joined and hiked with members of the
math and physics group where he met my mother. She had just received
her Doctorate in Economics and was working for a professor at the
university. In 1927 for a year Ernst worked as a teacher at the
famous boarding school, Schule Schloss Salem, in Salem, near Lake
Constance.  He then went back to Berlin University in 1928 so he
could begin studies on pedagogy and philosophy. Two years later, in
1930, he passed the state exams on higher education and they were
married.

\begin{figure}
\centerline{\includegraphics[height=8.0cm]{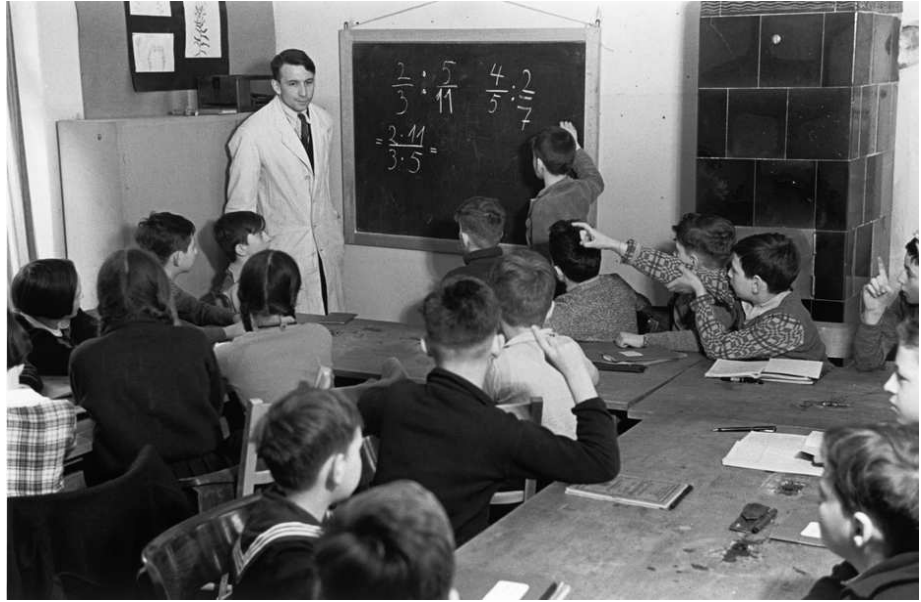}}
\caption{\label{caputh} Ernst Ising teaching one of his classes at
Landschulheim Caputh. $^\copyright${Herbert Sonnenfeld, Judisches
Museum Berlin.} }
\end{figure}

My parents moved to Strausberg where he had a teaching position and
my mother could take the train to Berlin. This lasted for two
wonderful years until April of 1933 when Jewish teachers were
removed from their positions. This was the start of ``12 years on a
tightrope" as my mother described it.

There followed a year of searching, including a very temporary job
at a school for emigrant children in Paris.

In 1934, he got a new job as a teacher for Jewish children at the
Judishes Landschulheim in Caputh, a few miles from Potsdam (see Fig.
\ref{caputh}). It was founded in 1931 by Gertrude Feiertag, who was
a known progressive social educationalist. Next-door was the
summerhouse of Albert Einstein. When Einstein permanently extended
his USA visit in 1932, the school rented his house to be used as
additional classrooms. This allowed the number of enrollees to
increase due to the fact that Jewish children were being expelled
from German public schools. Three years later my father took over
the headmaster position. But as one survivor said, ``the supposedly
safe island threatened to go under the brown sea at any time, and
the children and teachers knew that too".

\begin{figure}
\centerline{\includegraphics[height=8.0cm]{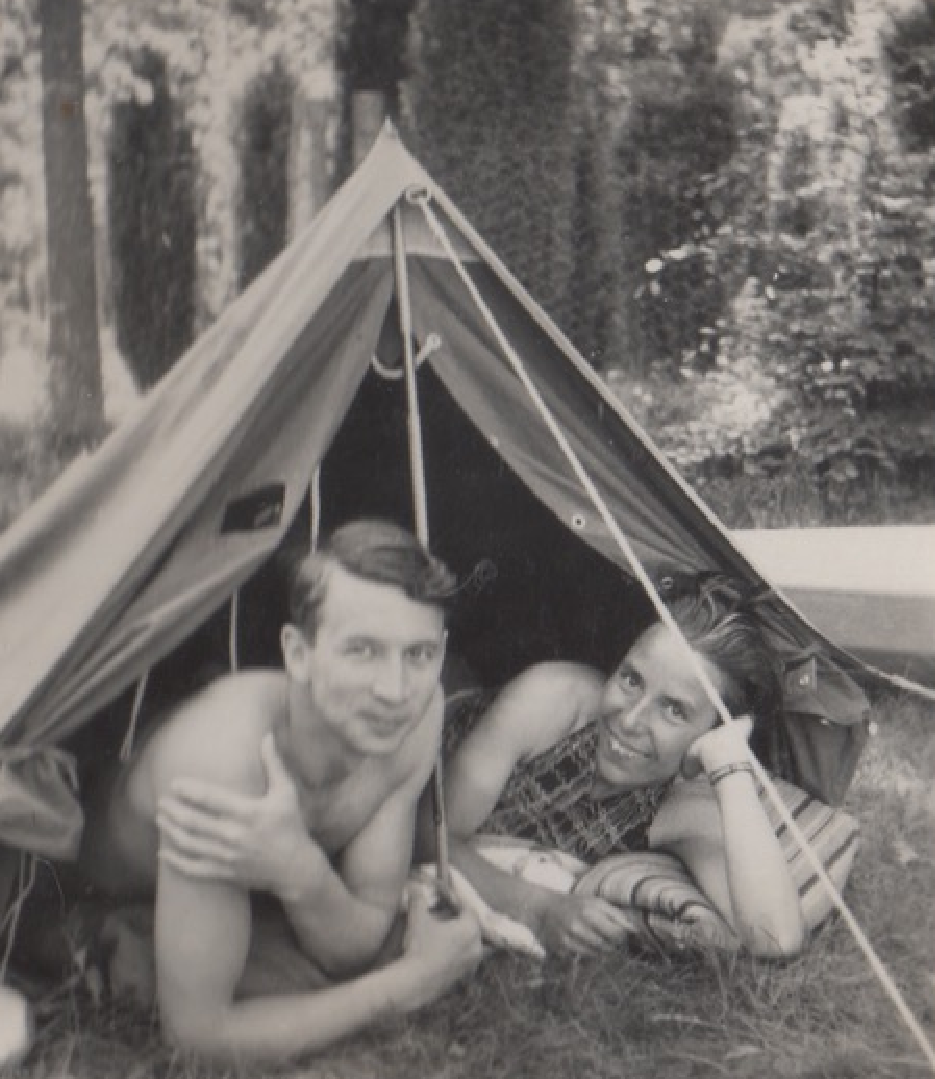}}
\caption{\label{camping} Ernst Ising and his wife Jane (Johanna)
Ehmer Ising during a camping trip down the Danube River in 1938.}
\end{figure}

While they were able to live near the campus by the relaxing Havel
River, it was possible to take a daily swim and take out their
Klepper foldboat, although the Nazi threat was constant. Once when
they thought my father was about to have a nervous breakdown, my
mother persuaded him to take a camping trip down the Danube River in
their foldboat (see Fig. \ref{camping}). At the end my father said
it was much better than a sanitarium.

On 10 November 1938, the school was destroyed, as part of
Kristalnacht, a program to get rid of the Jewish people in Germany.
The children had been prepared and were led in four groups through
the woods to transportation, home or safety. As one survivor put it,
``it was just like in the `Sound of Music'."

On 27 January 1939, he was taken by the Gestapo and interrogated for
four hours. He was only released after he promised that he and his
wife would leave Germany. They gained entry to the closed borders of
Luxembourg with the help of Dannie Heineman  (of the Dannie Heineman
physics prize) via his brother-in-law Hermann Busch and the Busch
Quartet. The quartet always gave two extra private performances in
Belgium, one for the Queen and one for Mr. Heineman. My parents had
planned to emigrate to the United States, but at that time the
quotas were full and they were forced to remain in Luxembourg where
I was born. Dannie Heineman had arranged for some 100 German-Jewish
families to occupy vacant hotels and to pay their room and board.
The Germans invaded on my father's 40th birthday. After the Germans
arrived Mr. Heineman made arrangements for one last payment of a
six-month allowance.

After that they survived in Luxembourg during the whole war by my
father doing mostly menial farm jobs. In between, there were ten
months of teaching Jewish children denied public schools in
Luxemburg City. Later there were several months of caring for sick
and old Jews who had not yet been deported from the Cinqfontaines
Monastery in northern Luxembourg. The Nazis had confiscated the
Monastery and were using it as a deportation center to send the Jews
to the camps. Near the end of the war, he was forced with other
mixed married Jewish men to help dismantle rails of the Maginot line
to be sent to the eastern front. He was left relatively unthreatened
as he had both a non-Jewish wife and an ``Aryan" son.

During this whole uncertain time they had acquired two used bicycles
and we were able to escape on long bike rides through the
countryside. Except for German control we were never in a war zone
until the end. On September 10, 1944 allied soldiers arrived in
Mersch and the horror was over. Later we escaped the Battle of the
Bulge by only ten miles.  Google indicates only 36 Jews survived in
Luxemburg.

\begin{figure}
\centerline{\includegraphics[height=8.0cm]{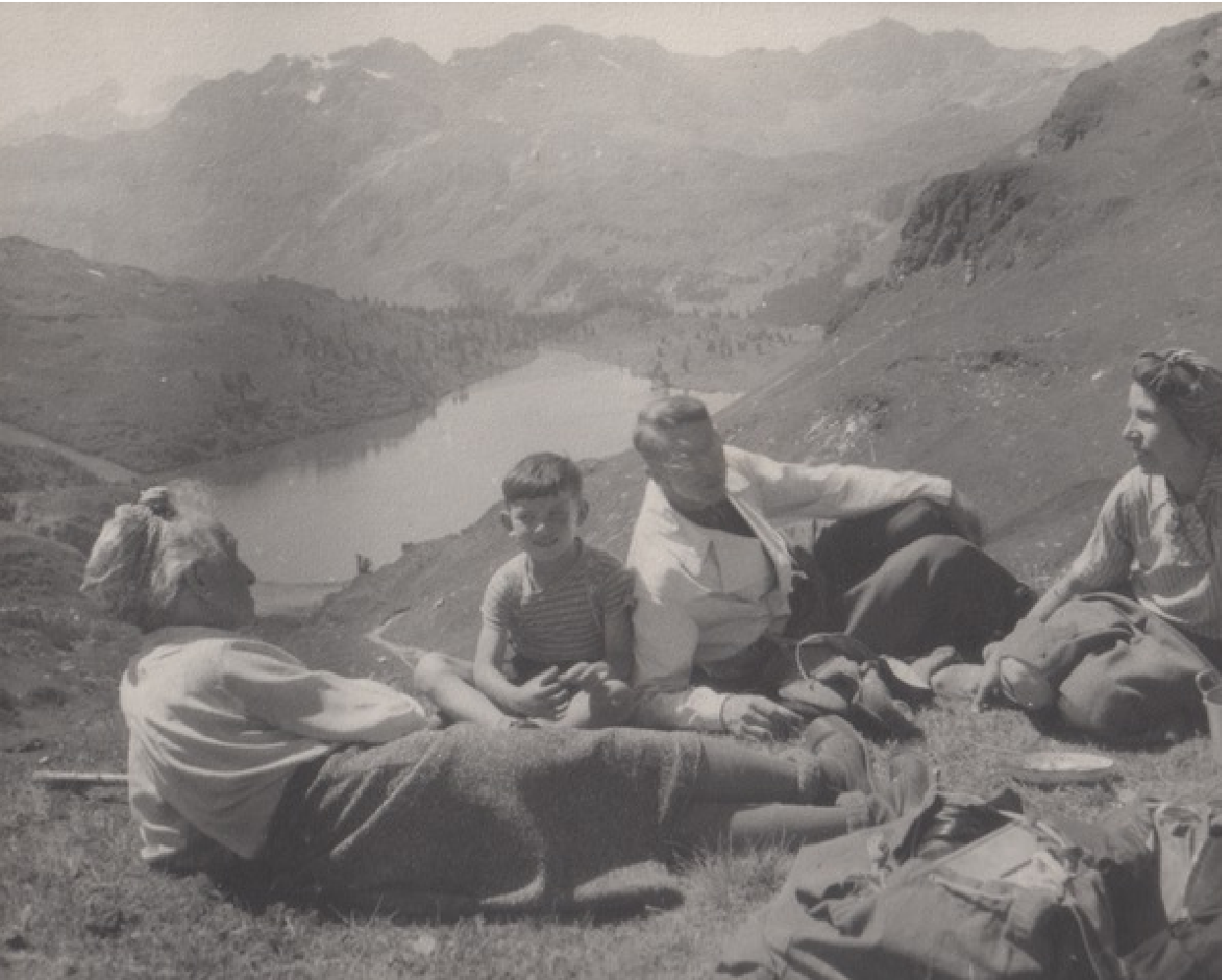}}
\caption{\label{switz} Thekla Ising, Tom Ising, Ernst Ising and
Johanna Ehmer Ising (from left to right) during their five-day hike
through the Swiss Alps (1946).}
\end{figure}

By 1946 we were able to take a month long vacation with my
grandparents who had been lucky enough to survive the war safely in
Basel, Switzerland. This included a five-day hike through the Alps
with my 71-year-old grandmother (see Fig. \ref{switz}).

It took over two years after the war ended for us to complete the
paper work necessary to enter the US. In April 1947, we finally
arrived in New York on the freighter  ``Lipscomb Lykes". That spring
my father went to a physics convention in Boston to get a job. There
he was asked for the first time if he was the ``Ising" of the Ising Model.

During that summer my parents found work at the Tapawingo Farm Camp
near Gouldsboro, PA. I was among other seven year olds who had Hanna
Tillich as our housemother. Our English improved tremendously.

That fall my father started as a teacher at the State Teacher's
College in Minot, North Dakota.  He had to make a very radical
change from teaching in a German high school nine years earlier to
teaching in an American college in English. The next year he became
a Physics Professor at Bradley University in Peoria, Illinois. His
wife Jane (Americanized from Hanna) also became a teacher at the
school. This was where they stayed. He retired in 1976. In 1953 we
were granted our US citizenships. He officially became Ernest and my
mother became Jane.

My parents soon made many lasting friendships. Every summer we, or
they, went on a significant trip, even driving up to Alaska one
year. They also took several trips to Europe and many other parts of
the world. On a lonely beach in Oregon the summer after my
graduation, they happened to meet one of my physics professors.  He
exclaimed that he was writing a chapter on the Ising model in his
new book.

My father passed away at home one day after his 98th birthday after
only five days in hospice.

\section{From Lenz {\em Umklappmagnets} via Ising's \\ Chain to Pauli's {\em Zweideutigkeit}}
\label{hist}

\begin{quotation}
{\em The most successful elaboration of technique in statistical
mechanics exists in connection with the Ising model.} (G. H. Wannier
1966 \cite{wannier})

{\em Starting around 1925, a change occurred: With the work of
Ising, statistical mechanics began to be used to describe the
behavior of many particles at once.} (L. P. Kadanoff
2013)\footnote{In \cite{kadanoff} Kadanoff cites together with
Ising's 1925 paper Brush's review \cite{brush} who made a similar
statement at the end of his paper.}
\end{quotation}

\subsection{Lenz paper from 1920}

Magnetism and especially ferromagnetism was a less understood
phenomenon at the beginning of the 20th century. Pierre Curie
discovered in 1895 that permanent magnets (ferromagnets) loose their
magnetization if they are heated above a certain temperature $T_C$,
now called Curie temperature \cite{pcurie0}. Curie recognized that
the behavior near the critical point in fluids and magnets seems to
be the same and introduced a kind of
universality\footnote{``Analogie entre la mani\`ere dont augmente
l'intensit\'e  d'aimantation d'un corps magn\'etique sous
l'influence de la temp\'erature et l'intensit\'e du champ, et la
mani\`ere dont augmente la densit\'e d'un fluide sous l'influence de
la temp\'erature et de la pression.'' \cite{pcurie0}} by pointing to
the ``analogy between the way in which the intensity of
magnetization of a magnetic body increases under the influence of
temperature and the intensity of the field, and the way in which the
density of a fluid increases under the influence of temperature and
of the pressure.''

Already in 1911 Niels Bohr and independently Hendrika Johanna van
Leeuwen discovered that magnetism is not a classical but a quantum
mechanical phenomenon. They proved\footnote{Bohr concluded in his
thesis: {\em a piece of metal in electric and thermal equilibrium
will not possess any magnetic properties whatever due to the
presence of free electrons} (see \cite{bohrthesis}, page 380).}:
``At any finite temperature, and in all finite applied electrical or
thermal fields, the net magnetization of a collection of electrons
in thermal equilibrium vanishes identically''
\cite{bohrthesis,leeuwen}.
 Bohr postulated
within his atomic model  that the planetary-like electron orbiting
around the atomic nucleus has a quantized angular momentum and
induces a magnetic moment. This condition allowed one to introduce
atomic (molecular) magnetic moments, which could respond to an
external field and to set up models for para- and diamagnetism. In
gases the magnets, according to the freedom of the atoms or
molecules, could be oriented in every direction. On the basis of
such assumptions Curie's law (the dependence of the susceptibility
$\chi$ with temperature $T$ as $\chi\sim c/T$) for paramagnets could
be derived.

In 1920 Lenz questioned the assumption of free rotation of the
elementary magnets in solids and suggested instead that they may
change their direction just turning around by 180 degrees
(Umklapp-Prozess) \cite{Lenz20}. He then derived Curie's law. In the
last paragraph of his short communication he suggested his two-state
model also for ferromagnets in order to explain the appearance of a
permanent magnetism at temperatures below $T_C$. He
says:\footnote{Nimmt man an, da{\ss} im ferromagnetischen K\"{o}rper
die potentielle Energie eines Atoms (Elementarmagnets) gegen\"{u}ber
seinen Nachbarn in der Nullage eiene andere ist als in der $\pi$
Lage, so entsteht eine nat\"{u}rliche zum Kristallzustand
geh\"{o}rige Gerichtetheit der Atome und daher spontane
Magnetisierung.} ``If one assumes that in ferromagnetic bodies the
potential energy of an atom (elementary magnet) with respect to its
neighbors is different in the null position and in the $\pi$
position, then there arises a natural directedness of the atom
corresponding to the crystal state, and hence a spontaneous
magnetization.'' (translation from \cite{Niss}$^{(a)}$).

In Weiss's domain model for ferromagnetism \cite{Weiss06} it is
the reaction of already ordered domains to a magnetic field which
leads to the Curie-Weiss susceptibility $\chi\sim c/(T-T_C)$,
whereas in Lenz's suggestion it is an unknown kind of non-magnetic
interaction between the two directions of the elementary magnets.

After Lenz took up the post of Chair of Theoretical Physics at the
University of Hamburg he was able to lead  a group of physicists and
young students to work on the project he suggested in his short
paper \cite{reich}. Einstein considered Lenz's papers on magnetism,
although published incompletely, as ``extremely important''
(\cite{reich} p. 93). The first to become involved in this project
was Ernst Ising, who was already a student in Hamburg when Lenz
became full professor. In 1922  Lenz proposed the problem outlined
in his 1920 paper for Ising's thesis, {\em Beitrag zur Theorie des
Ferro- und Paramagnetismus} \cite{isingthesis}.\footnote{{\em
Contribution to the theory of the ferro- and paramagnetism}.}

\begin{figure}
\centerline{\includegraphics[height=7.0cm]{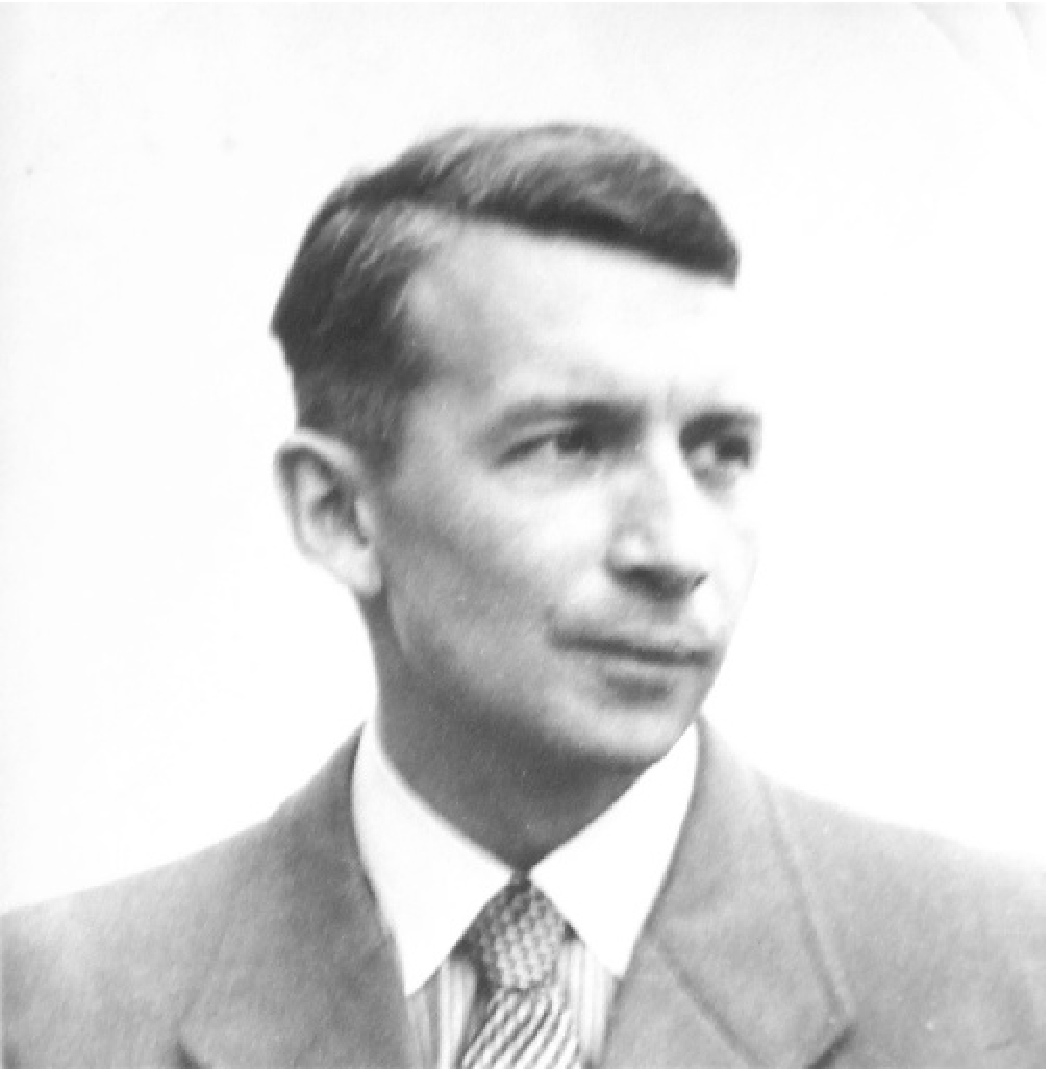}\hspace{2em}
\includegraphics[height=7.0cm]{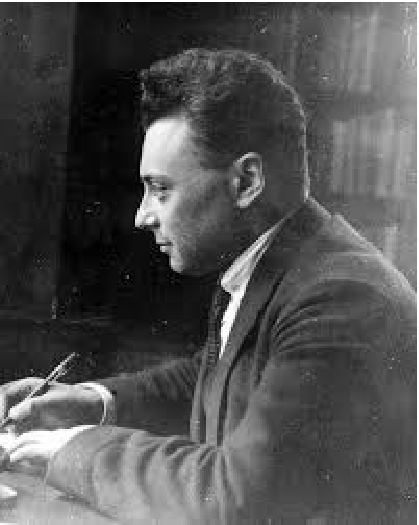}}
\caption{\label{ew}Ernst Ising and Wolfgang Pauli during the time in
Hamburg about 1925.}
\end{figure}

In May of the same year (1922) Lenz managed to obtain Wolfgang Pauli
as assistant.\footnote{In fact his title was ``wissenschaftlicher
Hilfsarbeiter''.} He was of the same age as Ising but already an
internationally well-known physicist. He came from G\"{o}ttingen
where he worked with Max Born \cite{enz} on problems with the
Bohr-Sommerfeld's atomic model and remained in Hamburg where he
submitted his Habilitation on 17 January 1924. He stayed in Hamburg
until March 1928 and moved to Z\"{u}rich when he got a Chair at the
Technical University. In 1923, during his stay at the institute of
Lenz, he visited for almost one year  Bohr's institute at
Copenhagen. During this time  Ising replaced Pauli until his return
at the end of September 1923 (see Fig. \ref{ew}).

\subsection{Ising's thesis and his 1925 publication}

The goal of Ising's task assigned by Lenz was to explain the
appearance of a ferromagnetic state in a three dimensional (3D)
solid. In fact this job was  a twofold one: First he had to set up
the model for the interaction of the elementary magnetic units,
which prefer alignment, a problem which belonged to the new and
undeveloped quantum mechanics. Second he then had to calculate
{\emph{analytically}} the macroscopic magnetization with the methods
of statistical mechanics.

Both tasks were  far too big to be solved in a thesis as the
development of quantum mechanics and statistical physics later
showed. The first problem one could say was answered in 1928 by
Heisenberg \cite{heisenberg} after the theory of quantum mechanics
proceeded far enough and the second in 1941 by Onsager
\cite{Onsager44} in 2D after special properties of the
model had been clarified by Kramers and Wannier and new methods of
calculating a partition function  had been found \cite{kramers}. The
3D problem remained analytically unsolved  until now (see
chapter \ref{Exact1} for more details).

Therefore Ising had to restrict himself for the first problem to
arguments for the model and for the second problem to reductions and
approximations. In order to come along with the first problem he
refers in the introduction  to the paper of E. A. Ewing
\cite{ewing}, ``where it was shown experimentally and theoretically,
that ferromagnetism is caused by a mutual interaction of the
elementary magnets.'' But the interaction is not thought to be the
well known interaction between dipoles. In fact ``no statement on the
nature of this force, which might be of electrical
nature\cite{schottky} can be made, but it is assumed that it decays
rapidly with the distance.'' It is interesting that these references
are missing in his 1925 publication. So Ising concludes a nearest
neighbor interaction is sufficient.  He further points out that this
is ``in crass disagreement'' with the hypothesis of  a  molecular
field. We now know that the critical behavior of systems with phase
transitions are described by mean field in dimensions high enough
otherwise it is an approximation or misleading as in the 1D case
here.

In order to attack the second task Ising restricted himself to the
1D case -- the famous Ising chain.\footnote{This problem is
reconsidered by Kramers and Wannier in the first paper of
\cite{kramers} in section 2 as an easy introduction to their new
method. In section 3 they explain: ``The reduction of the linear
chain problem can be described in a qualitative way as follows. It
is possible to build up a chain by repeating constantly one and the
same operation, namely {\emph{ adding another spin}} beyond the one
just placed previously'' (emphasis by the authors of this paper)
They explain, that the successful mathematical treatment is based on
one hand on the fact that no physical change takes place by this
procedure, if the chain is very long and on the other hand that the
state of the last added spin depends only upon the state of the
predecessor.''}  In the thesis the configurations on the chain are
displayed as vectors parallel or antiparallel to the  direction of
the chain.  This presentation  in the publication is replaced by the
short notation plus and minus restricting to orientations only
parallel or antiparallel to the direction of the chain.

\begin{figure}
\centerline{\includegraphics[height=3.0cm]{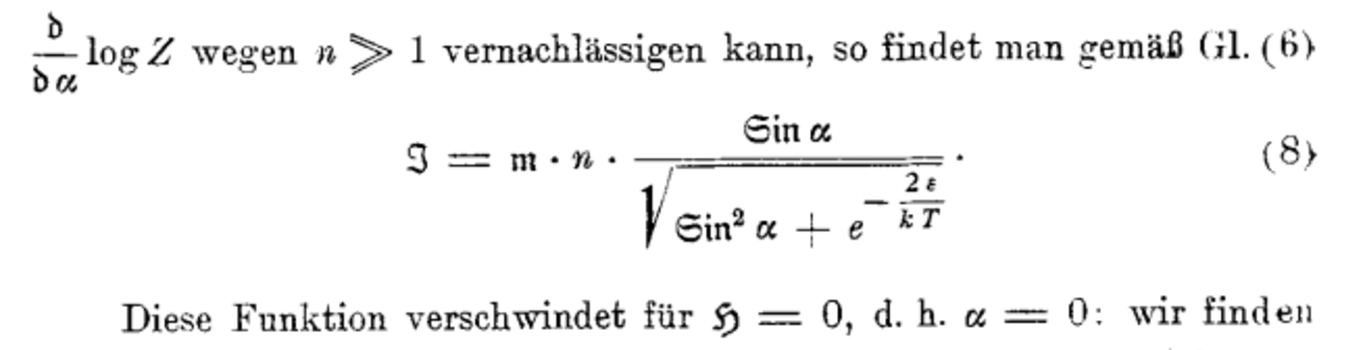}}
\caption{\label{mag25}Ernst Isings result for the magnetization of
the chain \cite{Ising25}.}
\end{figure}

The calculation follows the standard methods of equilibrium
statistical mechanics. Namely counting the configuration of
different energy in order to obtain the partition function and in a
next step the mean magnetization ${\cal J}$
\begin{equation}
{\cal J}=m\cdot n\cdot \frac{\sinh\alpha}{\sqrt
{\sinh^2\alpha+e^{-\frac{2\epsilon}{kT}}}}\, , \qquad
\alpha=\frac{mH}{kT}
\end{equation}
where $m$ is the elementary magnetic moment, $H$ is the external
field, $T$ the temperature, $k$ the Boltzmann constant and $n$ the
number of elements of the chain (see Fig. \ref{mag25}). Thus in zero
field no macroscopic magnetization arises at finite temperature.

Ising tried to generalize the model to higher
dimensions:\footnote{``Es ist ja denkbar, dass ein r\"{a}umliches
Modell, bei dem alle irgendwie benachbarten Elemente auf einander
wirken, die n\"{o}tige Stabilit\"{a}t mit sich bringt, um zu
verhindern, dass die Magnetisierungsintensit\"{a}t mit H
verschwindet. Doch es scheint in diesem Fall die Rechnung nicht
durchf\"{u}hrbar zu sein; jedenfalls ist es bisher nicht gelungen,
die Anordnungsm\"{o}glichkeiten geeignet zu sortieren und
abzuz\"{a}hlen.''} ``It is imaginable that a spatial model, in which
all elements that in some way are neighbors affect each other,
brings with it the necessary stability to prevent the magnetization
intensity to vanish with $H$. However, in that case the calculations
do not seem to be feasible; at any rate, so far it has not been
possible to sort and count the appropriate arrangement
possibilities.'' (translation from \cite{Niss}$^{(a)}$).   Indeed
this is not possible (so far) and one had to look for
approximations. He considered different kinds of arranging 1D
chains.  In the publication (section 3. ``The spatial model'') he
assumed the special limit where $n_1$ identical chains are spacially
arranged. He argues that differences in the configuration of the
interacting chains are energetically unfavorable. Therefore the
result is
\begin{equation}
{\cal J}=m\cdot n\cdot n_1\cdot\frac{\sinh n_1\alpha}{\sqrt {\sinh^2
n_1\alpha+e^{-\frac{2 n_1\epsilon}{kT}}}}
\end{equation}
and once again (although not surprising due to the approximation)
does not find a finite magnetization in zero magnetic field.

Based on these results in the thesis he concludes:\footnote{``Wenn
wir also nicht annehmen, wie dies P. Weiss tut, dass auch recht
entfernte Elemente einen Einfluss aufeinander aus\"{u}ben - und das
scheint uns auf keinen Fall zul\"{a}ssig zu sein - so gelangen wir
bei unseren Annahmen nicht zu einer Erkl\"{a}rung des
Ferromagnetismus. Es ist zu vermuten, dass diese Aussage auch
f\"{u}r ein r\"{a}umliches Modell zutrifft, bei dem nur Elemente der
n\"{a}heren Umgebung aufeinander wirken.''} ``So, if we do not
assume, as P. Weiss did, that also quite distant elements exert an
influence on each other – and this seems to us not to be allowed
under any circumstances – we do not succeed in explaining
ferromagnetism from our assumptions. It is to be expected that this
assertion also holds true for a spatial model in which only elements
in the nearby environment interact with each other'' (translation
from \cite{Niss}$^{(a)}$).

Ising finished his thesis in 1924 and published in 1925 a short
paper \cite{Ising25} with his results. There is not much known about
the contact between Ising and Pauli, but Brush \cite{brush} reports
a letter from Ising to him  where he stated ``...I discussed the
result of my paper widely with Professor Lenz and with Dr. Wolfgang
Pauli, who at that time was teaching in Hamburg. There was some
disappointment that the linear model did not show the expected
ferromagnetic properties...''. No further communication is  reported
apart from a letter from Pauli to Ising found by  Sigmund Kobe
\cite{kobe}$^{(d)}$, where Pauli informed Ising about his fate and
that of other colleagues in Hamburg after Ising left the institute
and which were also known by Ising.

\subsection{Pauli's struggle with the Bohr-Sommerfeld model of the atom}

The Bohr-Sommerfeld model of atoms was only partly successful in
explaining the experiments. It fails in cases where more than one
electron was present in the shell, but even in the case of one
electron discrepancies appeared.  The situation in the year 1923 is
explained by Land\'e in a short note \cite{lande}. He mentioned
that\footnote{Es zeigt sich n\"{a}mlich, da{\ss} bei Systemen aus
mehreren Elektronen nicht einmal die quantentheoretisch
station\"{a}ren Zust\"{a}nde und ihre adiabatischen \"{a}nderungen
mechanisch berechenbar sind.} ``It turns out that in systems with
more than one electron not even the quantum theoretical stationary
states and their adiabatic changes are mechanically calculable.'' He
notes as example the helium atom and adds:\footnote{Das zweite
besonders drastische Beispiel f\"{u}r das Versagen der mechanischen
Grundprinzipien auch in station\"{a}ren Quantenzust\"{a}nden gibt
die Multiplettstruktur und speziell der anomale Zeemanneffekt ...}
``The second particularly drastic example for the failure of the
mechanical basic principles also in stationary quantum states
illustrates the multiplet structure and especially the anomalous
Zeeman effect ...''.

After his stay in Kopenhagen Pauli gave his ``Antrittsvorlesung''
where he described the situation of the mechanical atomic theory. He
states:\footnote{``Der Inhalt dieser Vorlesung schien mir sehr
unbefriedigend, da das Problem des Abschlusses der Elektronenschalen
noch nicht weiter gekl\"{a}rt war. Das einzige was klar war, war,
da{\ss} eine engere Beziehung zwischen diesem Problem und der
Theorie der Multiplettstruktur bestehen mu{\ss}.''} ``The contents
of this lecture appeared very unsatisfactory to me, since the
problem of the closing of the electronic shells had been clarified
no further. The only thing that was clear was that a closer relation
of this problem to the theory of multiplet structure must exist.''
For another description of the desperate situation by 1924 see
\cite{kragh} page 125.

Another severe problem was the understanding of the periodic system
although Bohr constructed with help of an additional principle
(``Auf\-bau\-prin\-zip'') the structure of the shells in the
classical atomic model. Pauli tried to connect all these problems
and solve them with a new principle by postulating a fourth quantum
number for the electron and formulating his exclusion principle (the
name was given to it by Paul Dirac \cite{miller} p.59)  for the
electrons. He published those ideas  in 1925
\cite{pauli1925}$^{(a)}$ p. 385 and \cite{pauli1925}$^{(b)}$ p. 765
where he concludes:\footnote{``Die Dublettstruktur der
Alkalispektren sowie die Durchbrechung des Larmortheorems kommt
gem\"{a}{\ss} diesem Standpunkt durch eine eigent\"{u}mliche,
klassisch nicht beschreibbare Art von Zweideutigkeit der
quantentheoretischen Eigenschaften des Leuchtelektrons zustande.''}
``According to this point of view the doublet structure of the
alkali spectra, as also the piercing of the Larmor theorem, comes
about by a peculiar, classically not describable kind of
two-valuedness of the quantum mechanical properties of the valence
electron.'' (translation from \cite{enz} p. 107) The expression
nonclassic ``classically not describable'' was certified by later
development since Bohr was able to show that the spin could not be
measured by classically describable experiments \cite{paulinoble}.

The dramatic story of Pauli's struggle to increase the quantity of
quantum numbers from three to four is described by A. I. Miller
\cite{miller} (see also \cite{giulini,meyenn,jacobi}). Immediately
afterwards G. Uhlenbeck and  S. A. Goudsmit introduced for this
two-valuedness the concept of the spin for the electron
\cite{uhlenbeck}. Already in 1921 A. K. Compton discussed the
possibility that the electron possesses a magnetic moment as a
result of its spinning motion. A similar idea was formulated by
Ralph Kronig but never published.

After Ising had published his negative result it remained open in
the physical community if the higher dimensional cases would lead to
spontaneous magnetization or not.  Pauli communicated about
this question with Heisenberg (see \cite{hoddeson} p. 129 ff.) and
Heisenberg expressed his belief that if the number of nearest
neighbors (i.e. the dimension) is high enough one would succeed
finding ferromagnetism.

\subsection{The formulation of the Hamiltonian for the Ising model}

The quantum mechanical foundation of the interaction which might
lead to ferromagnetism was introduced in 1928 by Heisenberg
\cite{heisenberg}. It is known as the exchange interaction and is
due to the overlap of the wave function of neighboring atoms obeying
the exclusion principle. In this way the magnetic moments due to the
spin of the electron define the interaction. If the spins are
parallel the electrostatic energy is changed so that this
configuration is more favorable. He concluded in his
paper:\footnote{``(1) Das Kristallgitter mu{\ss} von solcher Art
sein, da{\ss} jedes Atom mindestens 8 Nachbarn hat. (2) Die
Hauptquantenzahl der f\"{u}r den Magnetismus verantwortlichen
Elektronen mu{\ss} $n\ge 3$ sein.''}``(1) The crystal lattice has to
be such, that each atom has at least 8 neighbors. (2) The main
quantum number of the electrons, which are responsible for the
magnetism has to be $n \ge 3$.''

In the year 1930 Pauli was invited to the Solvay conference  His
invited talk \cite{pauli1930}  gave a review of the status of the
theory concerning magnetism and its quantum mechanical nature (for a
short content of his talk see \cite{enz} page 220 ff.) Especially
interesting for  ferromagnetism is Section 5 of \cite{pauli1930}.
Here for the first time it is mentioned that the phase transition
could depend on dimensionality. He also mentions Ising's work in
connection with Heisenberg's work and its result for the magnetic
moment in molecular field theory
\begin{equation}
{\cal M}=N\mu_0\Big[1-C\Big(\frac{T}{\Theta}\Big)^{\frac{3}{2}}\Big]
\, .
\end{equation}
He states:\footnote{Ce r\'esultat est inter\'essant en liason avec
la discussion d'un mod\`ele semi-classique propos\'e par Ising.}
``There is in fact a very close relationship between the problem of
Ising and the one we have just treated''(\cite{pauli1930} p. 209).
Pauli's critical appreciation of Ising's model:\footnote{Dans le
calcul d'Ising, d\'evelopp\'e au point de vue de l'ancienne
th\'eorie des quanta, les composantes des $\sigma_i$
perpendiculaires \`a la direction du champ sont consid\'er\'ees
comme nulles, tandis que dans la nouvelle m\'ecanique cette
composante n'est pas commutable avec celle qui correspond \`a la
direction du champ.} ``In Ising's calculation developed from the
point of view of the old quantum mechanics, the components of
$\sigma_i$ that are perpendicular to the field are considered to be
zero, whereas in the new quantum theory these components do not
commute with the components in the direction of the field.''
(translation from \cite{Niss}$^{(a)}$ p. 291 slightly corrected) But
Pauli immediately suspects for the classical
variant\footnote{Malgr\'e cette diff\'erence, il est tr\`es
vraisemblable qu'une extension de la th\'eorie d'Ising au cas d'un
r\'eseau \`a trois dimensions donnerait du ferromagn\'etisme
m\^{e}me au point de vue classique.} ``Irrespective of this
difference, it is quite likely that an extension of the theory of
Ising to the case of a lattice of three dimensions would yield
ferromagnetism {\em even from the classical point of view}''
(emphasis by the authors of this paper).

\begin{figure}
\centerline{\hspace{-4.5cm}\includegraphics[height=6.0cm]{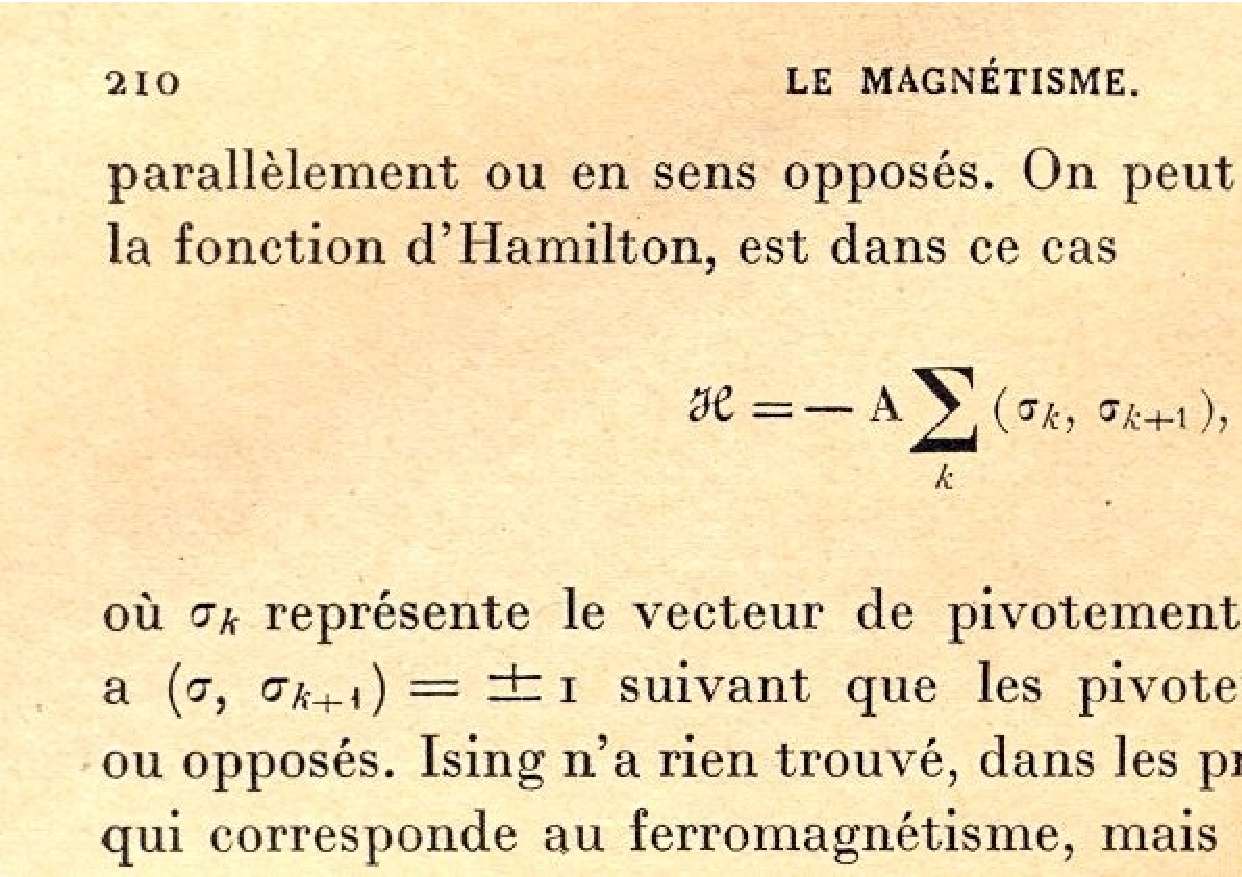}}
\caption{\label{pmodeli}Part of page 210 of Pauli's contribution to
the Solvay conference \cite{pauli1930}, where he presented the Ising
model in the form as it is known nowadays.}
\end{figure}

Thus it was Pauli himself who introduced the modern notation for the
Ising model \cite{pauli1930} p. 210, see Fig. \ref{pmodeli}
 \begin{equation}
H= -A\sum_k (\sigma_k,\sigma_{k+1})
\end{equation}
where $A$ gives the strength of the interaction of the spins on the
chain position $k$. Pauli pointed to the difference of the
properties of a quantum mechanical spin $\sigma_k$ and called the
``spin'' appearing in the Ising model a semiclassical spin.

This compact formulation of the Ising model includes already all the
aspects important for the following development: (1) the whole
system is described as an interacting many-particle system, (2) these
individual particles produce a specific collective behavior leading
eventually to phase transitions. The formulation also separates the
aspects the strength $A$ of interaction of the interacting units and
the properties of the units $\sigma_i$ themselves. $A$ is dependent
on the ferromagnet considered whereas the units are the same for the
whole group of ferromagnet. This reflects the important concept of
{\emph{universality}}, at least for ferromagnets, already introduced
1908 by Pierre Curie \cite{pcurie} within mean field theory and
comprising phase transitions in liquids and magnets. Future
developments like {\emph{scaling theory}} and {\emph{renormalization
group theory}} show that this universality concept goes beyond mean
field theory and the Ising model in three dimensions. Rather it
describes the critical behavior of a whole  universality class
containing liquids, magnets and  other physical systems of same
dimension, symmetry and type of short ranged interaction.

\subsection{Comments on Ising's result}
The usual explanation for the negative result for permanent
magnetization at finite temperature in the 1D case points to the
free energy. It consists of two parts, the internal energy and a
negative entropic term.  This entropic term favors disorder in the
1D case against macroscopic alignment. Another question was, if some
kind of long range interaction could change the result. Already from
the Curie-Weiss model it was known that taking into account the
interaction of all the spins by an effective field a phase
transition came about even in the 1D case. However an interaction
between two positions in the chain $i,j$ with a decay according to a
power law like $1/|i-j|^{1+\alpha}$ leads to a phase transition for
a sufficient weak decay, $\alpha<1$ \cite{luijten} (see also
\cite{mattis1d}).

Ising had to struggle with the configurations of the chain. A much
more elegant way, used mainly in textbooks, is to calculate the
partition function with the method of transfer matrices developed by
Kramers and Wannier \cite{kramers}.

Pauli's criticism that in fact quantum mechanics formulates a model
where the units are non-classical was taken up in the 1960's. It
turned out that (a) there are physical examples for which such a
model might be applicable and (b) numerical solutions of the problem
could be obtained \cite{bonner}. It also opened the new field of
quantum phase transitions.

\section{More on Exact Solutions}
\label{Exact1}


In the 1920's, the dominant theory for magnetism was that of Pierre
Weiss \cite{Weiss06}. This was based on the suggestion that
ferromagnets comprise  domains of parallel-aligned micromagnets.
Each micromagnet within a domain is supposed to experience an
effective magnetic field (the Weiss mean field) coming from its
neighboring magnetic moments. Each magnetic domain is then randomly
aligned, up to preferences induced by crystallographic symmetries.
Alternatives to Weiss's formulation include the Bragg-Williams
approximation \cite{BrWi34} as well as  Bethe-lattice models
\cite{Bethe}. The free energy coming from such mean-field approaches
is
\begin{equation}
 f(\beta,h) = \frac{qJm^2}{2} - \frac{1}{\beta}
 \ln{[2\cosh{h+J\beta q m}]},
\end{equation}
where $\beta = 1/kT$, $k$ is the Boltzmann factor, $T$ is the
temperature, $h=\beta H$ where $H$ is the strength of an external
field, $q$ is the coordination number (number of nearest neighbours
of a given site, e.g., $q=2d$ for a regular lattice of
dimensionality $d$), $J$ is the strength of the inter-site couplings
and $m$ is the mean-field magnetization. The model manifests a phase transition at
$h=0$ characterised by non-vanishing and vanishing values of $m$ on
either side of a critical temperature $T_c = qJ/k$. It also exhibits
discontinuity  in the specific heat  (the second temperature
derivative of the free energy) there.

Following the discovery the specific-heat anomaly of liquid helium
at temperatures of around 2.19K, Ehrenfest had introduced a
classification system for phase transitions \cite{Ehrenfest}. He
christened the anomaly the ``lambda point'' because of the shape of
the  experimentally obtained specific-heat curve. He argued that the
lambda point is a phase transition, even though it was dissimilar to
other known phase transitions in that it did not feature a latent
heat or change in volume. Ehrenfest had interpreted the lambda point
as a finite discontinuity
and he proposed to classify such  phase transitions as first- or
second-order depending on whether such a discontinuity in the first
or second derivative of the free energy. For a recent review of
Ehrenfest's scheme and a translation of his original paper
\cite{Ehrenfest}, see \cite{Sauer}. Thus the mean-field model
predicts a second-order phase transition in the original Ehrenfest
sense. Moreover this prediction holds for all dimensionalities.

In \cite{Ising25}, Ising explicitly highlights the difference
between his and Weiss's treatment in that only short-range,
nearest-neighbouring interactions are taken into account and the
orientation of each micromagnet restricted to  only two
possibilities. The solution for the free energy is
\begin{equation}
 f(\beta,h) = -\frac{1}{\beta}
 \ln{\left[{e^{\beta J}\cosh{\beta h} + \sqrt{e^{2\beta J}(\sinh{\beta h})^2+e^{-2\beta J}}}\right]},
\end{equation}
and the various thermodynamic functions are easily derived by
appropriate differentiation. As we have seen, unlike mean-field
theory, the model does not exhibit spontaneous magnetisation.

In 1936, however,  Rudolf Peierls showed that the model does
manifest ferromagnetism in two dimensions \cite{Peierls36} and  this
problem was investigated by Hendrik Kramers and Gregory Wannier in
1941 \cite{KrWa41}. They write in the introduction of their paper:
``The problem has a mechanical and a statistical aspect. On the
mechanical side we wish to improve our understanding of the
responsible coupling forces. On the statistical side we wish to
derive with certainty the thermal properties from a reasonable
accurate mechanical model. Both aspects have received extensive
attention. Quantum theory has explained satisfactorily the origin
and nature of the coupling forces. There are also several theories
available wich explain in terms of them the thermal behavior of
ferromagnets. Not one, however, applies just straight statistics to
the mechanical data. Generally some simplifying assumption is
introduced to facilitate the evaluation of the partition function.
It follows that the results obtained are not necessarily a
consequence of the mechanical model, but may well be due to the
statistical approximation.'' In their paper they then introduced the
transfer matrix concept and related the free energy of the Ising
model for high temperature to a conjugate Ising model at low
temperature. By using this relation they were able to calculate the
transition temperature of the 2D Ising model. They also developed
the  transfer matrix method and demonstrated  it by re-deriving the
results of Ising for the one dimensional chain. By their method they
reduced the calculation of the partition function to finding the
largest eigenvalue of a two by two matrix. For the two dimensional
Ising model the matrix turns out to be a square matrix of infinite
dimension and Kramers and Wannier could calculate the finite
transition temperature $T_c$. They showed  that the partition
function for the infinite system is related to the largest
eigenvalue of the matrix. They also discovered a symmetry in the
two-dimensional model in that its free energy at low temperature is
related to that at high temperature. The exact location for the
critical point of the model with square-lattice geometry is then
determined as the point which is invariant under this self-duality
transformation. It is given by $kT_c/J = 2/\ln{(1+\sqrt{2})} \approx
2.269185$. By way of comparison mean-field model theory gives
$kT_c/J = 4$ for $d=2$ and the Bethe approximation gives $2.88$.

Onsager solved the model for the square lattice in the absence of an
external field (i.e., with $h=0$) and famously announced his result
at the end of a talk by Wannier at the February 1942 meeting of the
New York academy of Sciences. He published the result in 1944 in
\cite{Onsager44}. The free energy for the infinite system in the
absence of an external field is
\begin{eqnarray}
-\beta f & = & \ln{2} + \frac{1}{8\pi^2} \int_0^{2\pi}{d\theta_1}{
\int_0^{2\pi}{d\theta_2}{ \ln{[\cosh{(2\beta J_1)}\cosh{(2\beta
J_2)} }}} \nonumber
\\
& & {{{ - \sinh{(2\beta J_1)}\cos{\theta_1} - \sinh{(2\beta
J_2)}\cos{\theta_2}]} }} ,
\end{eqnarray}
in which $J_1$ and $J_2$ are the coupling constants between spins in
the two different directions.

This was a milestone achievement in the history  of the Ising model
in that it was the first exact result for the model with a
finite-temperature phase transition and proved that these can be
captured by  statistical mechanics. Strictly speaking, the phase
transition was not of the Ehrenfest type --- it has a logarithmic
divergence instead of a discontinuity in the specific heat
\cite{KrWa41b}. Nowadays we consider Ehrenfest's classification
scheme as extended to include phase transitions with a divergence as
well as those with a discontinuity. The significance of Onsager's
achievement is reflected in a comment by Wolfgang Pauli to Hendrik
Casimir who had inquired about developments in theoretical physics
during the second World War: ``nothing much of interest has happened
except for Onsager's exact solution of the Two-Dimensional Ising
Model'' \cite{BhKh95}. Onsager's solution was simplified by Bruria
Kaufman in 1949 \cite{Kaufman} and Kaufman and Onsager determined
correlation functions in \cite{KaOn49}.

Onsager made another important announcement at the end of a talk by
L{\'{a}}szl{\'{o}} Tisza at Cornell University in 1948
\cite{BhKh95}. This time he stated that Kaufman and he had derived
the spontaneous magnetisation of the two-dimensional Ising model and
he wrote the formula on the blackboard. This was important because
  the non-vanishing of the magnetisation on one side of the
transition ($T<T_c$) and its vanishing on the other ($T>T_c$)
established the phenomenon as a genuine phase transition. Onsager
repeated the claim in May 1949 at a conference of the International
Union of Physics in Florence  after a talk by George Stanley
Rushbrooke \cite{BhKh95} but he and Kaufman did not publish the
derivation; the first to do so was  Chen-Ning Yang in 1952
\cite{Yang52}. Rodney Baxter recently reviewed how the
Kaufman-Onsager calculation was developed and added a draft paper
giving their result \cite{Baxter17}. That result is that the
spontaneous magnetisation behaves near the critical point as
$(T_c-T)^\beta$ with $\beta = 1/8$. This is different to the
mean-field result which is that $\beta = 1/2$. A deviation from mean
field exponents was already observed by   Verschaffelt
\cite{Verschaffelt00} in liquids around 1900 and corroborated by
further experimental material in different systems. With Kaufman in
1949 Onsager also derived the correlation function at the critical
point, and showed that it decays as $1/r^{1/4}$ \cite{KaOn49}.  In
1965, Alexander Patashinski and Valery Pokrovsky gave the
correlation-function exponential decay away from the critical point
\cite{PP}. The two-dimensional model has still not been solved in
the presence of an external field, apart from the
$c-$theorem approach of Zamolodchikov~\cite{Zamolodchikov} using
perturbed conformal field theories\cite{Henkel} where the model
turns out to be an example of integrable massive field theory.

In \cite{Niss}$^{(a)}$, Niss discusses the early years of the Ising
model within the context of quantum- and statistical-mechanical
models of magnetism. From the late 1940's and in the 1950's the
model was not believed to provide a good description of magnetic
materials due to its lack of physical realism \cite{Niss}$^{(b)}$.
The restriction of the interatomic forces to nearest-neighbouring
sites and the further restriction of spins of the Ising model to
only two orientations were considered to distance the model from
reality wherein the electron spin which can have any direction  in
three-dimensional space \cite{Brush83}.
It was thought that, at best, the Ising model may physically
represent anisotropic magnetic materials in which the two spin
directions were allowed or binary alloys with spins of each
orientation corresponding to one of the two types of atom in the
compound. It was also considered a model for lattice gas, in which
the presence or absence of a molecule at a point in space was
represented by one of the two spin orientations. But as a model of a
magnet, it was  considered lacking and its interest in this regard
was instead as a simplified model of phase transitions in general
that has the advantage of being  mathematical tractable
\cite{Niss}$^{(b)}$.

The group around Cyril Domb in King's College London did, however,
appreciate the physical importance of the Ising model
\cite{Niss}$^{(b)}$. They worked on different types of 2D lattices,
using geometries other than squares and  pioneered
 series-expansion approaches, a strategy also employed by the Rushbrooke group at the University of Newcastle.
They gained the crucial insight that  the critical exponents
describing the phase transition depend strongly on the
dimensionality of the system and less so on the geometry of the
lattices. This would later be explained by the notion of
universality (see section \ref{SecUniversality}). The  role of
dimensionality in the Ising model was therefore quite different to
that in mean-field theories, where it is unimportant for the
critical exponents \cite{Niss}$^{(b)}$.

Comparisons between series expansions and the exact solution in two
dimensions lent confidence that the approximate approach may be
applied to the three-dimensional version as well as to other models.
Indeed, and as discussed in \cite{Niss}$^{(c)}$, the Onsager
solution to the two-dimensional Ising model frequently played (and
continues to play) a role analogous to that traditionally played by
experiment in that hypotheses were tested against it. This includes
the scaling relations between the various critical exponents
describing continuous phase transitions. These establish that the
critical exponents are not all independent. The development of the
scaling relations by figures such as John Essam, Michael Fisher and
Benjamin Widom (and the related inequalities derived by Robert
Griffiths, Brian Josephson, Rushbrooke and others) were pivotal to
the development of more general theories of critical phenomena (for
a review, see e.g., \cite{Fisherrecoll}). They helped pave the way
for Widom's  hypothesis that the singular part of the free energy is
a homogeneous function  of its arguments. The explanation for
Widom's form was, in turn, given by Leo Kadanoff who ascribed the
singularity in the free energy to the occurrence  of large-scale
fluctuations in the system as the critical point is approached.
These fluctuations cause the correlation length to diverge and a
relation between temperature, field and length scales, a concept
captured by Kadanoff's  block-spin formulation  and ultimately by
Ken Wilson's renormalisation group. The renormalisation group forms
the foundation stone on which the entire modern theory of critical
phenomena is built and is of fundamental importance not just for
statistical physics but also for high-energy physics and any
physical system which can be viewed at different distance scales.
This also explained the crucial concept of universality, to use the
term coined by Kadanoff in 1971. This means that  critical exponents
are independent of many details of the Hamiltonian, and are
functions instead of the system  dimensionality, its internal
symmetries and  the range of interaction between its constituent
entities (spins) (see Section \ref{experiment} for experimental
verifications).

The three-dimensional Ising model has proved to be a far tougher
problem than its lower-dimensional counterparts and a solution
remains elusive, even in the absence of an external field It has a
status in statistical physics similar to that which Fermat's last
theorem occupied in mathematics, until proof of the latter by Andrew
Wiles in 1994; the problem is easily formulated but hard to solve.
Already in 1945  Wannier hoped that  an analytic solution was
imminent and both Onsager and Wolfgang Pauli are believed to have
attempted it in the 1950s \cite{Niss}$^{(b)}$. Other notable names
worked on a solution \cite{Maddox,Murray} ``and in the 1950s
physicists gradually concluded that a solution was not within
reach'' \cite{Niss}$^{(b)}$.

In 1986, Anders Rosengren reported an attempt to generalise
combinatorial considerations of the 2D nearest-neighbour model to
the three-dimensional simple cubic case. This led to the ``Rosengren
conjecture'' that the critical temperature for the 3D case is given
by $\tanh{(J/kT_c)} = (\sqrt{5}-2)\cos{(\pi/8)}$. This gives the
value $J/kT_c \approx 0.221\,658\,63$. Although this appears close
to the value $0.221\,654\,6(10)$ coming from simulational studies
\cite{BlLu95}, it is still over four standard deviations away. In
\cite{Fisher95}, Fisher showed that Rosengren's form comes from a
``critical polynomial''. A root of such a critical polynomial
delivers the critical point in the 1D and 2D cases and the hope was
that one could find the corresponding polynomial in the 3D case,
whose vanishing specifies its critical point. Fisher showed that
Rosengren's polynomial is a poor candidate; it does not mimic
desired features of the $d=2$ model, is not unique and the resulting
estimate for $T_c$ is not convincing.

The question of an exact solution of the 3D model has again come
under the spotlight recently and claims to have found the exact
exponents were given in Refs.\cite{Ka01,Zh07}. Rational values for
the critical exponents, including $\alpha = 0$ for the specific heat
have been given  with suggestions of  the existence of a
multiplicative logarithmic correction there \cite{Ka01,Zh07}. Such
claims are controversial because they are not in agreement with very
precise (and presumably accurate) approximations coming from a
variety of techniques including series expansions, renormalization
group, Monte Carlo simulations and experiment \cite{PeVi02}.
Additionally, although the values given in Refs.\cite{Ka01,Zh07}
obey the standard scaling relations, as they should, a logarithmic
term in the specific heat  would contradict  the scaling relations
for logarithmic corrections \cite{Kenna13}. The recent claims of
\cite{Zh07} and related papers \cite{March} were  criticised  in
Refs.\cite{FiPe16}.

Exact studies of the Ising model in low dimensions continue apace.
Boris Kastening recently presented a simplified version of Kaufman's
solution and extended it to various boundary conditions
\cite{Kastening,Kastening2}. Alfred Hucht exactly calculated the
partition function of the square lattice Ising model on the
rectangle with open boundary conditions for arbitrary system size
and temperature \cite{Hucht}. For the three-dimensional model, Sheer
El-Showk and collaborators produced a series of papers hoped to lead
to a solution of the conformal field theory for describing the three
dimensional Ising model at the critical temperature
\cite{ElShowk1,ElShowk2}. Their bounds are consistent with previous
estimates such as from renormalization group,   experiments and
Monte Carlo simulations. As they say in the final sentence of their
first paper: ``We have not yet solved the 3D Ising model, but we
have definitely cornered it'' \cite{ElShowk1}. In their second paper
they ask: ``Could it be that the critical 3D  Ising model is, after
all, exactly solvable?''  If not, El-Showk et al. at least have a
very efficient method to solve it numerically \cite{ElShowk2}.

Besides these exact results, a vast number of papers appear annually
which are related to the Ising model. Indeed, through the
renormalization group we now know that the validity of the Ising
model and its critical exponents extends far beyond anything that
could have been envisaged by Lenz or Ising in the 1920s, by Landau
in the 1930s or Onsager in the 1940s.

\section{Experimental Aspects of the Ising Model}
\label{experiment}
\subsection{Universality}\label{SecUniversality}
As discussed in the previous sections, a key theoretical concept of
critical phenomena which occur at second order phase transitions is
that of universality~\cite{Fisher67,KadanoffEtAl66}. According to
this concept, among the properties which describe critical
singularities in the neighborhood of a second order phase
transition, some exhibit a rather robust character, which means that
they do only depend on very general --- essential --- properties of
the system under interest. Other --- non-essential ---
characteristics are often called {\em details} in this context.
Among the essential characteristics, one usually mentions space
dimensionality, symmetries, range of interactions (see Section
\ref{Exact1}). The very nature of the interactions on the other
hand, such as whether they are of magnetic or of electric origin,
would they follow from classical or from quantum description of
matter, etc, is not essential. This robustness must also be
explained in deeper detail. Among the universal properties or
characteristics, the critical exponents which describe the leading
singularities of the thermodynamic quantities are probably the most
famous ones. Certain combinations of the critical amplitudes, these
numbers which appear in prefactors of the leading singularities,
also are universal. All these are just pure numbers, the set of
which defines a universality class. According to the universality
argument, let us assume that measurements are performed on some real
material which is expected to have the required symmetries to belong
to a given universality class. Then, extremely strong predictions
can be made for its critical properties. For example if a system is
expected to fall in the 2D Ising model universality class, the
critical exponent describing, let say its spontaneous magnetization,
has to be $1/8$. Not another number close to $0.125$, but exactly
$0.125$! And if it is not the case, then, the experiment is wrong!
This is  the incredibly strong predicting power of the theory of
critical phenomena. Of course, our statement that the experiment
would be wrong is exaggerated, and reality does not always simply
fits mathematical symmetries. Proving that a given material exhibits
the correct symmetries may be very challenging, but there are many
experimental situations in which the expected universal properties
can be measured. Thanks to universality again, although the real
material is often only approximately a representative of a given
universality class, deviations from the correct symmetry may appear
to be non-essential. We will illustrate below the concept of
universality with experiments performed on real materials which
belong to the Ising model universality class, either in 2D or in 3D.
There exist plenty of successful experiments and we will essentially
describe two of them which we consider particularly outstanding.

\subsection{Ising model behaviour in rare-earth materials}
The conditions to be fulfilled by real materials in order to be
quantitatively described by the Ising model are compelling. Magnetic
materials offer obvious candidates which are known to exhibit a rich
variety of phase transitions, with transition temperatures ranging
from very low to very high, as a result of the wide range of
variations of the magnetic interactions. We first have to understand
the behaviour of single magnetic ions in a crystalline environment
and two preliminary conditions are required. First the ground state
has to be a doublet separated energetically from the excited states
by a gap which is much larger than $k_BT_c$, where $T_c$ is the
transition temperature. Second, in order to keep the ground state
degeneracy, the operators involved in the spin-spin interactions
should all have vanishing matrix elements between the two Ising
states. For example, the exchange interaction $-J{\bf s}_i\cdot{\bf
s}_j$ transforms like a vector and as such, obeys the selection
rules $\Delta m=0,\pm 1$ where $m$ is the angular momentum
projection. Both conditions are often satisfied in compounds based
on rare-earth and one of the first materials which has been studied
in this context is the dysprosium ethyl sulfate,
Dy(C$_2$H$_5$SO$_4$)$_3$.9H$_2$O
~\cite{CookeEtAl59,CookeEtAl68a,CookeEtAl68b} with a doublet ground
state in the angular momentum state $|15/2,\pm9/2\rangle$ with weak
superposition of $|15/2,\mp3/2\rangle$  and $|15/2,\mp15/2\rangle$.
Local anisotropy axes  are furthermore parallel to the hexagonal
crystal axis. The system is thus well described  by a microscopic
Hamiltonian
\[  H=\frac 12\sum_{i,j} K_{ij}\sigma_{zi}\sigma_{zj} \]
where the sum extends over the pairs of spins $i$ and $j$,
presumably decaying with the distance among them. There is no
quantitative theory which would allow for a direct calculation of
the interaction strength $K_{ij}$, and these parameters have to be
obtained by the comparison between experimental results and
theoretical predictions of  thermodynamic quantities in regions of
the parameters where such theories are asymptotically exact, i.e.
when $T$ is either far above $T_c$ or far below $T_c$. This is for
example the case when the susceptibility is expanded in the moments
of the spin-spin interaction. Earlier studies then compared
experimental results with approximate theories: molecular field
models, cluster models, series expansions, etc, which, having no
adjustable parameters, were quite conclusive except maybe in the
very neighborhood of the transition.

Long power series expansions started to become available in the 1960's
and allowed for quantitative agreement in a wider range of
parameters, leading to the experimental determination of the 3D
Ising model critical exponents. The difficulty with fits to critical
point predictions is that the asymptotic range is generally very
narrow and limited by rounding effects which broaden the
singularities. These effects are described by corrections to
scaling, e.g.
\[ C(T,H=0)=A_\pm|t|^{-\alpha_\pm}(1+D_\pm|t|^{\omega_\pm})+B_\pm \]
with $t=(T-T_c)/T_c$, which require adjusting the experimental data
to non-linear fitting with in our example not less than 11
parameters (if we do not impose theoretical requirements like
$\alpha_+=\alpha_-$, etc)!

Similar studies then extended over half a century (extensive early
references can be found in the reviews
~\cite{JonghMiedema73,Stryjewski77,Wolff00}). Many experimental
problems were challenging. For example the presence of dipole-dipole
interactions lead to demagnetizing factors which result in a
sample-shape dependence, or to long-range interactions which modify
the upper critical dimension above which mean field exponents become
exact (the system under consideration is no longer in the Ising
universality class). Other phenomena which can be encountered
experimentally are field induced phase transitions (experimentally a
non-zero magnetic field is applied to promote one spin orientation
and single domain samples),  frustration (due to competing local
anisotropy axes), disorder (associated to the presence of vacancies
or defects). In spite of all these shortcomings which lead to rather
large differences between the model Hamiltonian and  the
experimental situation, the agreement between theory and experiment
is relatively unaffected, and this is a result of the extreme
robustness of universal quantities in the theory of critical
phenomena in general, and of the Ising model in particular which is
spectacularly exemplified below.


\subsection{A beautiful test of 2D Ising model universality}
Two-dimensional phase transitions may occur in very different
physical systems. The study of two-dimensional matter was initiated
in the XIXth century with molecular films of  non-soluble molecules
on liquid surfaces, and later with physisorbed atoms  on solid
surfaces. During decades however, investigators were not able to
observe experimentally the characteristics of two-dimensional
transitions, mainly because of the heterogeneity of the adsorbents
with multiple exposed crystal surfaces, defects, or chemisorbed
contaminants. In the 70's, lamellar solids, like graphite  appeared
well suited to such studies and nowadays, 2D adsorbed matter is the
subject of numerous works~\cite{Thomy81}. Reconstruction at crystal
surfaces also offer natural candidates to test experimentally
two-dimensional universality classes, e.g. the continuous structural
transition of Au(110) investigated through LEED experiments, which
appears to follow Onsager solution of the two-dimensional Ising
model~\cite{Campuzano85}.

But we will report here on wonderful experiments performed by C.H.
Back, Ch. W\"ursch, A. Vateriaus, U. Ramsperger, U. Maier and D.
Pescia~\cite{BackEtAl95}, where confirmation of a scaling behaviour
belonging to  the 2D Ising model universality class was shown to be
satisfied over 18 and 32 orders of magnitude in terms of the
properly scaled variables!

The experimental system consists in an atomic layer of ferromagnetic
iron deposited on a non-magnetic substrate made of single-crystal
W(110) surface, and provides a typical two-dimensional system. The
epitaxial growth guarantees crystalline order and avoids  disorder
(as much as possible). The fact that the system obeys Ising symmetry
(i.e. typically $\pm 1$ magnetization in normalized units) was
confirmed by the square shape of the hysteresis loop, measured by
magneto-optic Kerr effect. It also confirms the absence of domains
in the sample.
 In the vicinity of the critical point $t=(T/T_c-1)=0$, $H=0$, the temperature dependence of the spontaneous
magnetization  $M(t,H=0)$,  the critical isotherm $M(t=0,H)$ and the
zero-field susceptibilty $\chi(t,H=0)$ were measured, leading to the
corresponding critical exponents through $M(t,H=0)\sim (-t)^\beta$,
$M(t=0,H)\sim|H|^{1/\delta}$ and $\chi(t,H=0)\sim|t|^{-\gamma}$,
$\beta = 0.13\pm0.02$, $\delta=14\pm 5$ and $\gamma=1.74\pm0.05$.
This is a typical illustration of the possible experimental accuracy
which can be achieved, where 2D Ising expected exponents are $\beta
= 1/8$, $\delta =15$ and $\gamma=7/4$. Even more impressive is the
determination of the susceptibility amplitudes $\Gamma_\pm$ (via
expressions $\chi(t,H=0)\sim \Gamma_\pm|t|^{-\gamma}$) and their
ratio $\Gamma_+/\Gamma_-=40\pm 10$, where theory says that
$\Gamma_+/\Gamma_-=37.7$.

Testing universality can be pushed further. The scaling
hypothesis~\cite{Widom65,DombHunter65,PatashinskiiProkovskii66,Kadanoff66,PatashinskiiProkovskii}
states that thermodynamic functions can be written in the vicinity
of the critical point as generalized  homogenous functions, e.g.
\[ M(t,H)=b^{-\beta/\nu} \tilde m(b^{1/\nu}t,b^{\beta\delta/\nu}H) \]
where $b$ is an arbitrary scaling factor. Fixing $b=1/M^{\nu/\beta}$
above yields $\tilde m(t/M^{1/\beta},H/M^{\delta})=1$, which then
allows to write the parametric equation of state in terms of
rescaled variables,
\[
H/M^{\delta} = f(t/M^{1/\beta}).
\]
The experiment of Back {\em et al.} reported this rescaled equation
of state fitted to theoretical results~\cite{GauntDomb70} over 18
orders of magnitude in the variable $t/M^{1/\beta}$ and almost 32
orders of magnitude in $H/M^{\delta}$!

\begin{figure}
\centerline{\includegraphics[height=8.0cm]{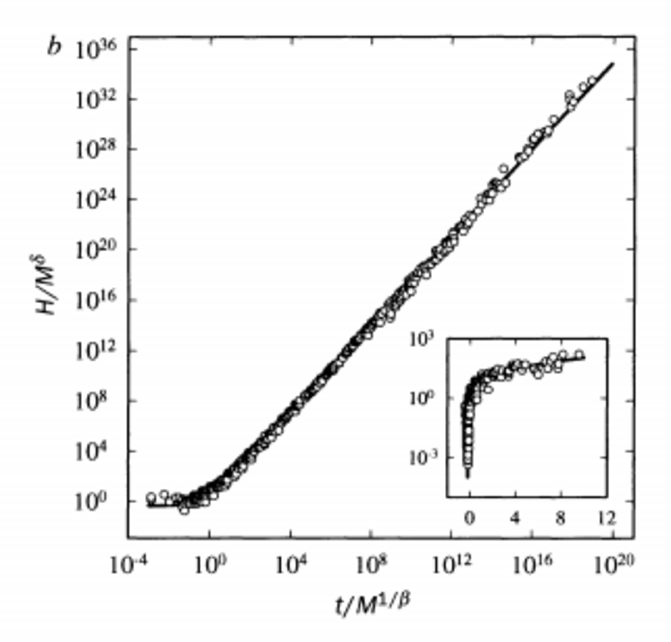}}
\caption{\label{Exp1} Universal plot from \cite{BackEtAl95}.}
\end{figure}

 This might be considered as a
real achievement and an incredible success of the theoretical
prediction, which even raises the opposite question: how is it that
all experimental imperfections, inhomogeneities which break
translational symmetry,  non-localized local magnetic moments (Fe is
a broad-band metallic ferromagnet when Ising Hamiltonian is written
in terms of localized ones), non-perfect uniaxial local symmetry and
possibly other sources of discrepancy do not destroy the 2D Ising
model universality class. Although there cannot be a simple answer
to such questions, some of these effects are understood within the
frame of universality. Disorder for example can be shown not to
change (up to logarithmic corrections) the 2D Ising model
universality class~\cite{Harris74}, and the experimental evidences
reported in \cite{BackEtAl95} support a scenario where all
imperfections mentioned eventually prove to be non-essential (we say
irrelevant in the renormalization group
language~\cite{MaBook,Wilson74}).

\section{Simple Model of Complex Systems}
\label{exot}

\hspace{10em}{\em But let your communication be, Yea, yea; Nay,
nay:}\\
\hspace*{10em}{\em   for whatsoever is more than these cometh of
evil.}\\
\hspace*{25em} (Matthew 5:37)

It is stated sometimes, that although Ising model is not realistic,
its success to a large extent is caused by the fact that it allows
analytic treatment (see e.g. \cite{Niss} and discussions therein).
In this sense, it belongs to the ``narrow class of models which are
balanced (precariously!) between realism and solubility''
\cite{Kac71}.
This is certainly true, but our belief is that there
is another -- even more important -- reason for the tremendous
success of the Ising model.
The simplicity of the model not only
enables its analytic treatment, it also singles out an essential
feature: binarity, i.e. representation of something as a pair of
binary oppositions (cf. Umklappmagnets in section \ref{hist}).
It is
this feature that enables  a much wider set of applications of the model.
Moreover, and as we will see
below, this feature has aided  the  exportation of very notions of physics to other fields,
giving rise to science of complex systems \cite{Holovatch17}.
Indeed, the model tailored by the ``usual
procedure of separating the phenomena till one deals with simple
elementary facts'' (as Ising himself noted on a different occasion
in his other paper \cite{Ising50}) singles out the notion of
binarity and enables analytic treatment. This is corroborated by the
remark in \cite{lasGuevas13} that such an approach in condensed
matter theory ``consists of building a model of the system which is
simple enough to handle, but rich enough to capture the relevant
properties. These simplifications give rise, among others, to
classical spin models. A paradigmatic example is the Ising model
[26], originally devised to study magnetism." In turn, this enables
one to apply the model in almost all fields where binarity plays a
core role. Sometimes this role is not obvious from the very
beginning and this is the skill of researchers to find a subtle
connection between the cause and the consequence.

In the epigraph to this chapter we have chosen probably one of the
oldest written references suggesting binary opposition
\cite{Stauffer00}. Indeed, binary variables (plus or minus one, up
or down, filled or empty, active or passive) are ubiquitously used
in describing various processes occurring in nature and in human
society.  Quantitative descriptions of such processes, on the
one hand, allow us to apply methods developed in one field to
another, and on the other hand, it triggers a search for
similarities between very different phenomena and ordering them into
different classes. It also fosters transfer of knowledge from one
branch to another.
 As we will see from several examples given below,
applications of the Ising model for quantitative descriptions and
understandings of different phenomena of physical, chemical,
biological, or social nature, as well as  its application in humanities is based on
the fundamental fact that actually the very essence of these
phenomena is hidden in their {\em statistical nature}. In the
so-called agent based modeling that lies in the core of such
descriptions, one considers a whole system as a set of agents (individuals in
social systems; spins in magnets) that are capable of
autonomous behaviour. Usually, an agent has a well-defined internal
state and interacts with other agents. Allowing such agents to be in
one of two possible states leads to the Ising model description.

Currently, there are numerous applications of the Ising model to explain
chemical or biological phenomena. An amount of studies  and their
success lead to the situation when e.g. such typically biological
phenomena as dynamics of pattern formation in neural networks
\cite{Rojas96,Schaap05} or protein folding \cite{Munos01} became
conventional and well-established fields of physics. The Ising model is
being successfully used to explain properties of living organisms on
all scales. Just to give some examples, on a molecular and cellular
scale, it is adapted to the analysis of complex genetic models with
several genetic effects and with interaction, or epistasis, between
the genes (see \cite{Majewski01} and references therein) and serves
as a framework for phase transitions in multicellular environments
\cite{Weber16}. On the other extreme, at the scale of ecosystems,
it explains how a critical transition can emerge directly from the
dynamics of ecological populations \cite{Noble15}. In ecology,
long-range synchronization of oscillations in spatial populations
may  elevate extinction risk. Therefore, such phenomena may signal
an impending catastrophe.

The above examples of Ising-model applications, although outside
physics still concern systems that traditionally belong to natural
sciences. As a next step, let us illustrate how it is applied in
social sciences, where an important topic is to understand the
social dynamics of a community, e.g. its transition from an initial
disordered state to a configuration that displays at least partial
order
\cite{Durlauf99,Stauffer04,Stauffer08,Galam08,Castellano09,Galam12,Ball03,Ball04}.
Inspired by an idea to exploit binarity in social choice, T.C.
Schelling has suggested a model to describe racial segregation in
cities \cite{Schelling71}. There, in particular, special attention
is paid to analysis of the relation between individual and
collective states: ``But evidently analysis of `tipping' phenomena
wherever it occurs - in neighborhoods, jobs, restaurants,
universities or voting blocs--and whether it involves blacks and
whites, men and women, French-speaking and English-speaking,
officers and enlisted men, young and old, faculty and students, or
any other dichotomy, requires explicit attention to the dynamic
relationship between individual behavior and collective results''
\cite{Schelling71}. Although the phenomenon of interest in the above
example is rather the phase separation and not an onset of a phase,
an analogy with Ising model is obvious but it has not been
recognized in the original paper. Only later the similarities
between phase separation into domains in the Ising model at $T=0$
and residential segregation in the Schelling model were recognized
and the equivalent of the temperature $T$ was introduced into the
Schelling model \cite{Stauffer07}.

When the authors of \cite{Galam82} identified binarity of states of
social agents to describe the  phenomenon of strikes, the analogy
with the Ising model was apparent. As Serge Galam recalls in his
book: ``we developed the idea of using an Ising ferromagnetic system
to describe the collective state of an assembly of agents, each
being in either one of two distinct individual states, that of
working or striking. This produces two collective ordered states: a
working state versus a striking state. The ferromagnetic coupling
between agents was motivated by the social fact that people have the
tendency to reproduce the leading choice of their neighbors, in
particular in conflicting situations. We thus implemented the first
application of the Ising model to describe the global state of a
firm...'' \cite{Galam12}. Currently, analysis of opinion dynamics
widely exploits agent based modeling with agents being in discrete
binary states. The most widely used in this context models are the
voter model \cite{Clifford73,Holley75}, majority rule models
\cite{Galam02}, the Sznajd model
\cite{Sznajd-Weron00,Sznajd-Weron05} and other models based on a
social impact theory \cite{Latane81} and its extensions
\cite{Schweitzer00,Schweitzer03}. A detailed review of these and
other models may be found in \cite{Castellano09}.

\begin{figure}
\centerline{\includegraphics[height=4.0cm]{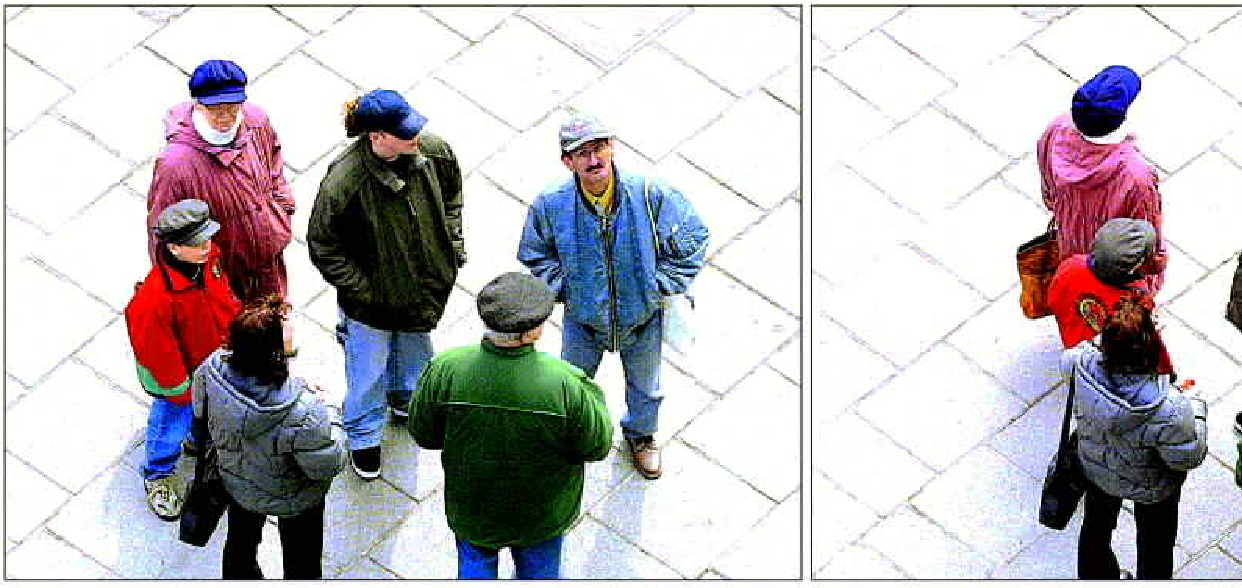}}
\caption{\label{soc1}People, similar to magnets, may experience
symmetry breaking. At the beginning all of them look in different
directions (low order, high symmetry). Then somebody shouts from the
other side and all start staring in the same direction (high order,
low symmetry). And this is in spite the fact that only one of them
has heard the call: curiosity serves as an interaction between the
people. (Illustration and caption is taken from the mass media
article about phase transitions: {\em Der Standard}, 02.04.2002,
Austria).}
\end{figure}

Concepts of phase transition theory, and, more specifically, the
Ising model are being also actively used in the field of economics and
financial markets, explaining, in particular,  statistical
properties that are common to a wide range of financial assets
\cite{Mantegna99,Bouchaud00}. In modeling financial markets, the
agents are identified with spin variables which can take specific
values depending on agents decisions:  the +1 spin as a buyer and -1
as a seller
\cite{Bornholdt01,Kaizoji00,Kaizoji02,Krawiecki02,Zhou07,Sornette06,Dvorak12}.
Considering also the case when an agent may stay inactive ($S=0$)
leads to further generalization
\cite{Iori99,Takaishi05,Sato07,Sieczka08,Denys13}. Such approaches
allow to study inherent features observed in collective behaviour of
financial markets: herding,  bubbles or crashes and to reproduce
main statistical observations of the real-world markets such as
fat-tailed  distribution  of  returns  or volatility clustering.

In Social Sciences, we can also mention elegant applications of the
Ising model to Natural Language Processing, via the ability of
magnetic models of statistical physics to extract the essential
information contained in texts. Documents are represented as sets of
interacting magnetic units (words), and a textual energy is defined
as an indicator of information relevance which allows automatic
abstract production, information retrieval, document classification
and thematic segmentation. The compression of a sentence appears as
the ground state of the chain of terms and variants are produced by
thermal fluctuations \cite{Silvia}.

In almost every example given above the Ising model was used to shed
light on the  behaviour of systems composed of many interacting
agents, which display collective behavior that does not follow
trivially from the behaviors of the individual parts. Such systems
are currently known as {\em complex systems} \cite{complex_system}.
Their inherent features incorporate self-organization, emergence of
new functionalities, extreme sensitiveness to small variations in
the initial conditions, power laws governing their statistics
(fat-tail behaviour) \cite{Mitzenmacher03,Newman05,Simkin11}. Their
systematic study gave rise to  complex system science: the field of
knowledge that is actively developed and shaped
nowadays.\footnote{{It is worth mentioning here words of Wolfgang
Pauli from his letter to Herman Levin Goldschmidt (Feb. 19, 1949)
\cite{vonMeyenn}: ``It seems to me as a philosophical layman that
the task of philosophy consists in generalizing the emerging
insights of current physics – that is, all its essential elements –
in such a way that it can be applied to fields more general than
physics. Such an achievement would, in turn, enrich the individual
disciplines and prepare future developments.''}} Usually,
quantitative description of such systems is achieved by considering
agents located on the nodes of a graph called complex network
\cite{Albert02,Dorogovtsev03,Newman06}. Linking between graph nodes
corresponds to the interaction between the agents under analysis.
For social systems it corresponds to social interactions, for
ecological systems it may reflect predator-pray relation between
species, for transportation systems it correspond to transportation
links, etc. In this sense, treating the Ising model on complex
networks has various applications in complex system science. Of
special importance are the so-called small-world \cite{Watts98} and
scale-free \cite{Barabasi99} networks. The first are characterized
by small characteristic sizes (usually, their typical size $\ell$
logarithmically grows with number of nodes $N$: $\ell \sim \ln N$).
The second are characterized by the power-law decay of the  node
degree distribution $P(k)\sim k^{-\lambda}$. Many important natural
and man-made networks are small world and scale-free. {Examples are
given by the internet, world-wide web, some transportation,
biological, social networks \cite{Albert02,Dorogovtsev03,Newman06}.
Properties of the Ising model on such types of networks essentially
differ from its properties on $d$-dimensional lattice. The
scale-free networks with slowly decaying node-degree distribution
(fat-tailed distributions with small $\lambda$) are highly
inhomogeneous. It appears, that the decay exponent $\lambda$ plays a
role in some sense similar to that of dimensionality $d$: Ising
model on a scale-free network with $\lambda\leq 3$ is ordered for
any finite temperature $T$ whereas it has a finite $T$ second order
phase transition for $\lambda>3$. Moreover, the basic concept of
universality is revised: the critical exponents attain
$\lambda$-dependency in the region $3< \lambda <5$
\cite{Leone02,Dorogovtsev02,vonFerber11}  and the logarithmic
corrections to scaling appear at $\lambda=5$
\cite{Palchykov10,Kenna13}.

These and many more unusual features of the Ising model on complex
networks are currently well established by different approaches (see
\cite{Dorogovtsev08} for a review) and recently revisited by
Lee-Yang-Fisher zeros analysis \cite{Krasnytska15,Krasnytska16}.

There are at least two lessons one can learn from the short account
given in this chapter. Indeed, exploiting Ising archetype in
agent-based modeling of various complex systems of chemical,
biological, social, economical origin gives a possibility to
quantify them and to understand some of the mechanisms of their
behaviour. {In this sense the model enables one to single out
universal common features of different systems. However, more than
this: it would be too trivial to reduce behaviour of these systems
just to a single archetype no matter how powerful and general the
archetype is.} Along with universality in behaviour of many-agent
interacting systems, they are characterized by system-specific
diversity. Subtle changes in their parameters may lead to crucial
changes in their global behaviour: this is another inherent feature
of complex systems. In their description, the Ising model plays a role
of the main `course', however these are the spices which make the
whole dish tasty.

We have already mentioned Ising's paper \cite{Ising50} at the
beginning of this chapter.  There, discussing Goethe's approach to
analyze nature he says the following: ``...his approach to
science was that of an artist who thought he could conceive the
secrets of nature in all their complexity... He was convinced that
translation into language of mathematics  was distortion of
reality...''. Contrary to Goethe's believe,  nowadays  Ising-like
models completed by ideas from complex system science come into play
as simple models on the way to ``conceive the secrets of nature in
all their complexity''.

\end{document}